\pdfoutput=1

\documentclass[amsmath, amssymb, aps, longbibliography, prl, superscriptaddress,
reprint,
 amsmath,amssymb,
 aps,
]{revtex4-2}

\usepackage[dvipsnames]{xcolor}
\usepackage{graphicx}
\usepackage{dcolumn}
\usepackage{bm}

\usepackage{siunitx}

\newcommand{\Freiburg}{Institute of Physics, University of  Freiburg, Hermann-Herder-Straße 3, 79104 Freiburg, Germany}
\newcommand{\FMF}{Freiburg Materials Research Center, Universit\"at  Freiburg, Stefan-Meier-Straße 21, 79104 Freiburg, Germany}
\newcommand{\MBI}{Max Born Institute for Nonlinear Optics and Short Pulse Spectroscopy, 12489 Berlin, Germany}
\newcommand{\XFEL}{European XFEL GmbH, Holzkoppel 4, 22869 Schenefeld, Germany}
\newcommand{\FLASH}{ Deutsches Elektronen-Synchrotron DESY, Notkestr. 85, 22607 Hamburg, Germany}
\newcommand{\Rostock}{Institute of Physics, University of Rostock, Albert-Einstein-Straße 23-24, 18059 Rostock, Germany}
\newcommand{\ETH}{Laboratory for Solid State Physics, ETH Zurich, 8093 Zurich, Switzerland}
\newcommand{\TU}{Institut für Optik und Atomare Physik, Technische Universit\"at Berlin, Hardenbergstraße 36, 10623 Berlin, Germany}
\newcommand{\Hamburg}{Department of Physics, Universit\"at Hamburg, Luruper Chaussee 149, 22761 Hamburg, Germany}
\newcommand{\IWM}{Fraunhofer IWM, MikroTribologie Centrum $\mu$TC, W\"ohlerstraße 11, 79108 Freiburg, Germany}
\newcommand{\fig}{Fig.\,} 
\newcommand{\RostockChem}{Institute of Chemistry, University of Rostock, Albert-Einstein-Straße 3a, 18059 Rostock, Germany}
\newcommand{\RostockLLM}{Department Life, Light and Matter, University of Rostock, Albert-Einstein-Straße 25, 18059 Rostock, Germany}

\begin{document}

\preprint{APS/123-QED}

\title{Melting, bubble-like expansion and explosion of superheated plasmonic nanoparticles}

\author{Simon Dold} 
    \homepage{contributed equally}
\affiliation{\Freiburg}
\affiliation{\XFEL}
\author{Thomas Reichenbach} 
    \homepage{contributed equally}
\affiliation{\Freiburg}
\affiliation{\IWM}
\author{Alessandro Colombo}
    \homepage{contributed equally}
\affiliation{\ETH}
\author{Jakob Jordan}\affiliation{\TU}
\author{Ingo Barke}\affiliation{\Rostock}\affiliation{\RostockLLM}
\author{Patrick Behrens}\affiliation{\TU}
\author{Nils Bernhardt}\affiliation{\TU}
\author{Jonathan Correa}\affiliation{\FLASH}
\author{Stefan D\"usterer}\affiliation{\FLASH}
\author{Benjamin Erk}\affiliation{\FLASH}
\author{Thomas Fennel}\affiliation{\Rostock}\affiliation{\RostockLLM}
\author{Linos Hecht}\affiliation{\ETH}
\author{Andrea Heilrath}\affiliation{\TU}
\author{Robert Irsig}\affiliation{\Rostock}
\author{Norman Iwe}\affiliation{\Rostock}
\author{Patrice Kolb}\affiliation{\ETH}
\author{Bj\"orn Kruse}\affiliation{\Rostock}
\author{Bruno Langbehn}\affiliation{\TU}
\author{Bastian Manschwetus}\affiliation{\FLASH}
\author{Philipp Marienhagen}\affiliation{\RostockChem}
\author{Franklin Martinez}\affiliation{\Rostock}
\author{Karl-Heinz Meiwes-Broer}\affiliation{\Rostock}\affiliation{\RostockLLM}
\author{Kevin Oldenburg}\affiliation{\Rostock}\affiliation{\RostockLLM}
\author{Christopher Passow}\affiliation{\FLASH}
\author{Christian Peltz}\affiliation{\Rostock}
\author{Mario Sauppe}\affiliation{\TU}\affiliation{\ETH}
\author{Fabian Seel}\affiliation{\TU}
\author{Rico Mayro P. Tanyag}\affiliation{\TU}
\author{Rolf Treusch}\affiliation{\FLASH}
\author{Anatoli Ulmer}\affiliation{\TU}\affiliation{\Hamburg}
\author{Saida Walz}\affiliation{\TU}

\author{Michael Moseler} \affiliation{\Freiburg} \affiliation{\IWM}
\author{Thomas M\"oller}\affiliation{\TU}



\author{Daniela Rupp}
 \email{ruppda@phys.ethz.ch}
\affiliation{\ETH}\affiliation{\MBI}
\author{Bernd von Issendorff}
 \email{bernd.von.issendorff@uni-freiburg.de}
\affiliation{\Freiburg}\affiliation{\FMF}

\date{\today}

\begin{abstract}
We report on time-resolved coherent diffraction imaging of gas-phase silver nanoparticles, strongly heated via their plasmon resonance. The x-ray diffraction images reveal a broad range of phenomena for different excitation strengths, from simple melting over strong cavitation to explosive  disintegration. Molecular dynamics simulations fully reproduce this behavior and show that the heating induces rather similar trajectories through the phase diagram in all cases, with the very different outcomes being due only to whether and where the stability limit of the metastable superheated liquid is crossed.

\end{abstract}

\maketitle
Condensed matter, when heated slowly, undergoes the familiar phase transitions from solid to liquid to gas. Extremely fast heating, on the other hand, results in additional phenomena, such as strong overheating of solids or liquids or even a change in the chemical bonding itself due to strongly excited electrons \cite{recoules2006effect,lindenberg_atomic-scale_2005,ernstorfer2009formation,medvedev2020nonthermal}. Many aspects of matter under such extreme conditions, like the coupling between a very hot electron gas and the ionic system, are not yet fully understood \cite{mo2021ultrafast,lee2021investigation}.
Isolated nanoscale particles have been identified as well-controlled test objects for the study of highly excited matter \cite{saalmann2006mechanisms, fennel_laser-driven_2010, nishiyama2019ultrafast}, especially in combination with single-shot coherent diffraction imaging (CDI) using intense x-ray free-electron laser pulses \cite{neutze2012time, marchesini_coherent_2003, raines_three-dimensional_2010, loh_fractal_2012, bostedt2012ultrafast}, which permits to follow the dynamics with high spatiotemporal resolution  \cite{gorkhover2016femtosecond}. Up to now, most time-resolved studies concentrated on rare gas particles in free flight, excited by strong laser fields \cite{gorkhover2016femtosecond, fluckiger2016time, nishiyama2019ultrafast, bacellar2022anisotropic, langbehn2022diffraction}, and a few other systems like silicon dioxide particles \cite{peltz2022few}. Time-resolved diffraction studies on metal nanoparticles were mainly performed using particles supported on surfaces, leading to the observation of vibrational excitation, melting, or disintegration \cite{ihm2019direct, clark2015imaging, clark2013ultrafast, von2016watching, jung_inducing_2021, sung2021single, wu2014microscopic, shin_ultrafast_2023}. 
However, an unambiguous analysis of the particle dynamics free from the spurious and mostly unknown influence of the support is only possible in the gas phase. 

In this work, we study silver nanoparticles heated in free flight by excitation of their plasmon resonance with moderately intense picosecond 
laser pulses.
Such  particles have been shown to form well-defined faceted crystalline structures at lower temperatures \cite{reinhard1997size,barke20153d, colombo2023three}, distinctly different from the round shape of a liquid droplet, which should facilitate the observation of melting. Furthermore, they exhibit a strong Mie plasmon resonance in the near UV \cite{kreibig1985optical,el2001some}, which permits to excite them using fairly weak laser fields \cite{el2001some,klar1998surface,link2000laser,petrova2007time,hartland2011optical}.
  This excitation leads to a rather uniform heating of the particle via very fast formation of a hot electron gas and a subsequent transfer of the energy to the nuclear degrees of freedom on a time scale of several picoseconds \cite{hartland2011optical,link2003optical,hartland2017s,voisin2001ultrafast,beane2018ultrafast,hoeing_time-resolved_2023}.  Employing pump-probe CDI, we determine the morphology of the particles as a function of delay time. Accompanying molecular dynamics simulations of the process provide comprehensive insight into the origin of the observed phenomena. The strength of a decompression wave following the almost isochoric heating of the particles turns out to be a decisive parameter for the dynamics.

The experiment was performed at the free-electron laser (FEL) FLASH \cite{ackermann2007operation} using the CAMP endstation \cite{erk2018camp}.  Silver nanoparticles with diameters in the range of \SIrange[range-phrase=-,range-units=single]{50}{200}{nm} were produced by a magnetron sputter gas aggregation source  \cite{haberland1991new,haberland1994filling}, operated with a mixture of xenon and argon. The beam of neutral nanoparticles traversed a differential pumping stage before entering the main chamber. Here, the particles were intercepted by \SI{20}{\micro J} FEL pulses with about \SI{70}{fs} pulse duration at \SI{5.1}{\nm} wavelength, focused to a spot size of \SI{10}{\micro m}. Photons scattered from the particles were collected by a pnCCD detector  \cite{struder2010large} at a distance of \SI{70}{\mm} from the interaction region at a repetition rate of \SI{10}{Hz}.
Heating of the particles was achieved by \SI{400}{\nm} laser pulses synchronized with the FEL pulses 
and overlaid in a near-collinear geometry. They were produced by a frequency doubled Ti:Sapphire femtosecond laser system and stretched to a duration of about \SI{1}{ps} by propagating through 100 mm of fused silica. We note that due to the time scale  of the electron-phonon heat transfer of several picoseconds, shorter pulses would not increase the temporal resolution, but would just lead to  unwanted strong fields effects. The \SI{50}{\micro\joule} pulses were weakly focused to a spot size of roughly \SI{70}{\um} diameter in the interaction region. Inhomogeneities in the laser beam profile did not permit a precise UV laser intensity determination; we estimate a maximal value of \SI[per-mode=symbol, tight-spacing=true]{1e12}{\watt\per\cm\squared}.

\begin{figure}[b]
\includegraphics[width=0.44\textwidth]{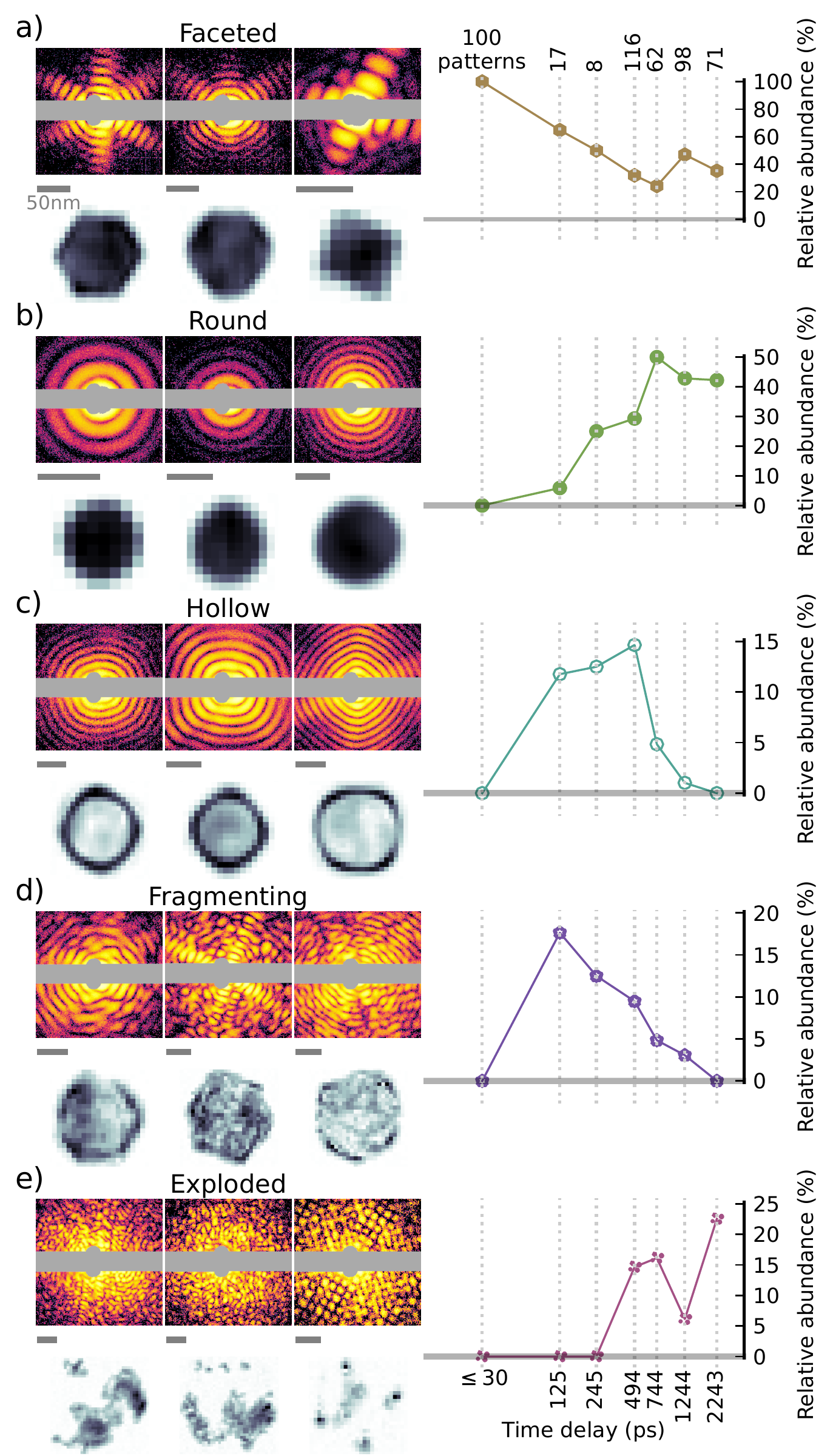}
\caption{\label{fig:observations} Characteristic classes of images. Three examples are provided for each class, with the experimental diffraction patterns in logarithmic color scale and the corresponding reconstructions in linear gray scale (object sizes indicated by \SI{50}{\nm} bars). In the right column, respective relative abundances of each class as function of the time-delay between pump laser and FEL pulses are given, with the absolute number of patterns at each delay at the top. 
}
\end{figure}

In \fig\ref{fig:observations} (a)-(e) examples of scattering images recorded with the pump pulse present are shown, along with real-space reconstructions obtained by iterative phase retrieval reconstruction of the small angle part of the images \cite{colombo2023three} (see the Supplemental Material \cite{suppl} for details).
Five different classes can be identified, based on the real-space features of the samples. The \emph{faceted} class in \fig\ref{fig:observations}(a) contains polyhedra with well-defined facets, which appear not to be strongly affected by the pump laser, either because the images were recorded at negative or small positive delay times, or because the particles have interacted only with a low-intensity part of the pump laser profile. \fig\ref{fig:observations}(b) 
represents examples of the \emph{round} class - these are mostly spherical particles with homogeneous density, most probably liquid droplets melted due to the heating.  \fig\ref{fig:observations}(c) shows the most spectacular class, \emph{hollow} particles with a rounded outer surface and a large, close to spherical cavity inside. In some cases the remaining shell has a thickness of only about \SI{10}{\percent} of its diameter.
The  \emph{fragmenting} class (\fig\ref{fig:observations}(d)) refers to particles with a very inhomogenous density, indicating disintegration probably due to fairly strong excitation.
The \emph{exploded} class (\fig\ref{fig:observations}(e)) includes all patterns that correspond to unconnected fragments of the original particle, which leads to speckle-like diffraction patterns. These obviously are particles which completely disintegrated after the interaction with the pump laser. Interestingly, rather symmetric speckle patterns are sometimes observed (like in the third image in this row), corresponding to a small number of large fragments in a rather regular arrangement.

\begin{figure*}[t]
\includegraphics[width=0.9\textwidth]{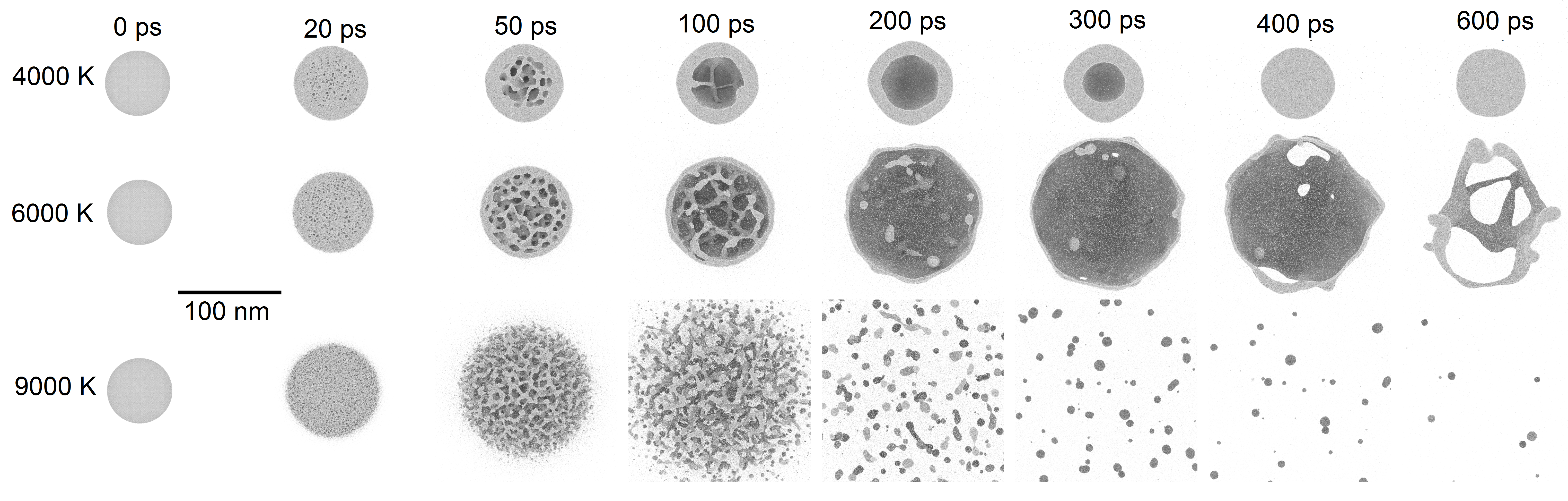}
\caption{\label{fig:md} MD simulations of strongly heated silver particles with \num[tight-spacing=true]{8e6}\,atoms. The snapshots show cross sections for different times, for three different heat bath temperatures. See the Supplemental Material \cite{suppl} for videos of the dynamics.}
\end{figure*}

The relative abundances of these five classes as a function of the time delay between the pump laser and the FEL pulse are plotted in the right column of \fig\ref{fig:observations}. The total number of identifiable diffraction images recorded at each time delay is given on the top, showing the overall low hitrate in this experiment. For negative and small positive time delays up to \SI{30}{\ps} (here grouped together in a single data point) only faceted structures were observed. Other morphologies start to appear  at a delay of \SI{125}{\ps}. The abundance of \emph{faceted} samples drops to a level of about \SI{40}{\percent} within the first few hundreds of picoseconds. The abundance of \emph{round} specimen exhibits the opposite behavior, rising up to \SI{40}{\percent} in the first few hundreds of picoseconds. \emph{Exploded} patterns appear even later, at around \SI{500}{\ps}, and tend to become more and more abundant towards longer time delays.
The other two classes appear only transiently. The  contribution of the \emph{fragmenting} class is significant only between \SI{125}{\ps} and \SI{744}{\ps}; a similar behavior is seen for the \emph{hollow} class, with the maximum abundance apparently shifted to slightly longer delays.

In order to better understand the observed dynamics, the temporal evolution of strongly heated silver particles was simulated by classical molecular dynamics (MD), employing the embedded-atom-method interaction potential of Sheng et al.\@ \cite{sheng2011highly}. Particles with \num[tight-spacing=true]{8e6}
atoms were simulated starting from initially spherical particles (R=\SI{31}{\nm}), a shape which can be most easily compared to analytical models. In order to mimic the heating process by electron-phonon-coupling, the particles were thermalized for a duration of \SI{10}{\ps} using a Langevin thermostat with a given temperature in a range between \SI{2000}{\kelvin} and \SI{9000}{\kelvin}, using a relaxation time of \SI{5}{\ps} (see the Supplemental Material \cite{suppl} for details). This would in principle allow the atomic system to reach a temperature of \SI{86}{\percent} of the heat bath temperature. However, the actual temperatures at the end of the heating interval are lower due to cooling by evaporation, melting and expansion during the heating. 
Results of the MD simulations for three different heat bath temperatures are shown in \fig\ref{fig:md}. Snapshots of the nanoparticles are depicted at different times after the start of the heating process. In the Supplemental Material \cite{suppl}, videos of the dynamics are provided. In general, the simulation results are in line with those of earlier studies  \cite{wu2014microscopic,jiang2021molecular,gan2022ultrafast,castro-palacio2020hollow}. In the case of a heat bath temperature of \SI{4000}{\kelvin} (first row), small voids start to form in the inner part of the sphere briefly before \SI{20}{\ps}. These voids grow and coalesce into a single large cavity in the center of the particle within \SI{200}{\ps}. This central cavity then contracts again, leading to a round liquid droplet at a delay time of 400 ps.
For the case of \SI{6000}{\kelvin} (second row), the initial behaviour is similar, but the void formation now happens even closer to the particle surface. As before, the voids merge to form a single central cavity within \SI{200}{ps}, but with a substantially thinner outer shell. In this case, the surface tension is not strong enough to reverse the expansion of the bubble; it further expands and bursts in less than a nanosecond, forming rather large, initially tubular and eventually round fragments.
The last case represents the dynamics of the system for a heat bath temperature of \SI{9000}{\kelvin}. Here, the formation of voids happens in the whole volume, even at the surface of the sample. This leads to a direct and violent disintegration of the sample within the first \SI{100}{\ps}, with fragments spreading at significant velocities.

These results are in good agreement with the experimentally observed time evolution (\fig\ref{fig:observations}).
As mentioned, the \emph{faceted} particles surviving up to long time delays (\fig\ref{fig:observations}(a)) most probably have only weakly interacted with the pump laser.
The appearance of round droplets around \SI{500}{\ps} is compatible with still rather weakly excited particles, which just melt, or with slightly more strongly excited ones, where a central cavity appears and vanishes again, like in the simulation for \SI{4000}{\kelvin} heat bath temperature (\fig\ref{fig:md}). The observation of large bubbles only between \SI{125}{\ps} and \SI{744}{\ps} is congruent with the simulations at both \SI{4000}{\kelvin} and \SI{6000}{\kelvin}, where the bubble vanishes again at larger times, either by contraction or by fragmentation. 
The rapid increase of \emph{fragmenting} samples already at \SI{125}{\ps} can be assigned to early, fairly violent void formation, somewhere between the behavior simulated for \SI{4000}{\kelvin} and \SI{6000}{\kelvin}. The still significant presence of these samples up to around \SI{1}{\ns} hints at processes more in line with the \SI{6000}{\kelvin} simulation, indicating bursting bubbles. Finally, the \emph{exploded} samples can stem as well either from early violent fragmentation or a late bursting of a bubble - especially the third example in \fig\ref{fig:observations}(e) hints at such a case, because of the small number of larger fragments detected. When comparing measured and simulated time dependencies, one should keep in mind that the size of the particles in the simulation is on the lower end of the size distribution of particles observed in the experiment; for the larger particles the dynamics can be expected to be slower than in the simulation. 

\begin{figure}[t]
\includegraphics[width=0.85\columnwidth]{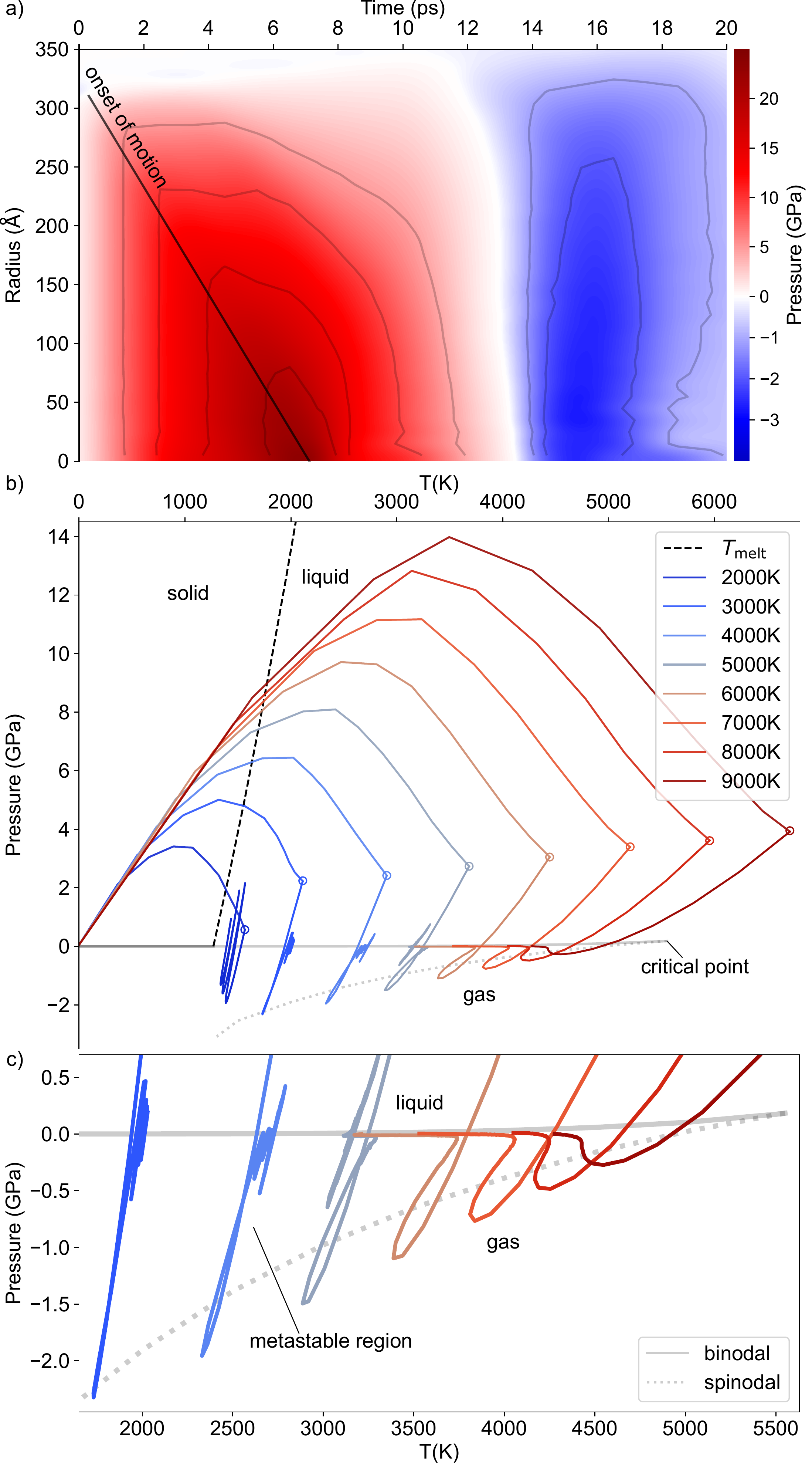}
\caption{\label{fig:pressureAndPhaseSpace}Evaluation of molecular dynamics simulations. a) Radial pressure distribution at different times, for a heat bath temperature of \SI{5000}{\kelvin}. The solid line indicates the propagating boundary between material at rest and in  motion. b) Evolution of the system within the phase diagram of silver; the circles indicate the end of the heating process at 10 ps. c) Detailed view of the metastable region.}
\end{figure}

The question arises why the single large cavity is always formed at the center of the particle. This can be understood with the help of the simulation results, as shown in \fig\ref{fig:pressureAndPhaseSpace}. Rapid uniform heating of a particle leads to the buildup of a high pressure, which will eventually lead to an overall expansion of the particle. This expansion, however, does not occur via a simple breathing mode motion, but instead in form of a decompression wave: as an acceleration of the material requires a pressure gradient, the motion starts at the particle surface, with the boundary between moving material and material at rest propagating into the particle with the velocity of sound. \cite{paltauf2003photomechanical,wu2014microscopic,fahdiran2018laser,jiang2021molecular}. 
\fig\ref{fig:pressureAndPhaseSpace}(a) shows the pressure within the particle during the heating phase and the following \SI{10}{\ps} as a function of radial position for the case of a \SI{5000}{\kelvin} heat bath. One can see that a pressure of more than \SI{20}{\giga\pascal} is reached in the particle center after \SI{6}{\ps} of heating. The motion of the material sets in immediately when the heating is started, with the boundary between moving and non-moving layers propagating towards the particle center. The position of this boundary is indicated by the black line in \fig\ref{fig:pressureAndPhaseSpace}(a) (see the Supplemental Material for radial profiles of the velocity distributions \cite{suppl}). It  reaches the particle center at about \SI{7}{\ps}, which gives a velocity of sound of about \SI[per-mode=symbol]{4400}{\metre\per\second} under these conditions, slightly higher than the room temperature value of \SI[per-mode=symbol]{3650}{\metre\per\second} \cite{CRC2017}. When this happens, all of the atoms are moving outwards, which leads to a strong pressure drop within the particle. This pressure drop is slightly less abrupt than might be expected due to the heating continuing until \SI{10}{\ps}; nevertheless, \SI{14}{\ps} after the start of the heating, a negative pressure of about \SI{-2.5}{\giga\pascal} is reached (\fig\ref{fig:pressureAndPhaseSpace}(a)). As further discussed below, at such a negative pressure cavitation sets in immediately, leading to the many voids as seen in \fig\ref{fig:md} at \SI{20}{\ps} for \SI{4000}{\kelvin} and \SI{6000}{\kelvin} heat bath temperature. The voids grow and merge into a larger cavity, surrounded by material which continues to move outward as it keeps the momentum gained in the first phase of the expansion. We want to stress that it is not the vapor pressure which drives the growing of this bubble, but rather the outward movement of its shell. Therefore, if this motion is not too fast, it can be reversed by surface tension, like in the \SI{4000}{\kelvin} case where the large central void starts to shrink again at about \SI{200}{\ps} and has completely vanished at \SI{400}{ps}. For higher bath temperatures, the initial shell velocity is so high that the surface tension could reverse the motion only at much later stages of the expansion. Before that happens, however, any small hole formed in the thinning liquid layer leads to the rupture of the bubble \cite{taylor1959dynamics,lhuissier2012bursting}.

At even higher heat bath temperature the particle rips apart immediately. Nevertheless, the underlying mechanism is quite similar to the case of the cavitation process discussed before, as can be best seen by comparing the system trajectories through the phase diagram of silver as shown in \fig\ref{fig:pressureAndPhaseSpace}(b) and \ref{fig:pressureAndPhaseSpace}(c). In \fig\ref{fig:pressureAndPhaseSpace}(b) the state of silver is shown as a function of pressure and temperature; in \fig\ref{fig:pressureAndPhaseSpace}(c) the region most relevant for the experiment is magnified. This phase diagram has been obtained from bulk silver MD simulations using the same interaction potential as for the silver particle simulations (see the Supplemental Material \cite{suppl}). In it, the liquid and gaseous region are separated by the so-called binodal, which towards higher temperatures ends in the critical point. Below the binodal, matter should always be in the gaseous state. Nevertheless, the phase transition here is kinetically hindered by seed bubble formation; there is a critical bubble size, which has to be overcome, and which increases the closer the system is to the binodal. Only at a second boundary, the so-called spinodal, the liquid becomes unstable and homogeneous evaporation sets in \cite{paltauf2003photomechanical,vogel2003mechanisms,wu2014microscopic}. Thus, in the region between the binodal and the spinodal, the liquid can be overheated, staying in a metastable liquid state, while this is not possible anymore beyond the spinodal. The evolution of the average temperature and pressure within the particles with time is indicated for eight different heat bath temperatures. 
One can observe a strong initial increase of temperature and pressure upon heating. This in fact moves the system away from the liquid-gas boundary, therefore somewhat stabilizing the liquid. Due to the particle expansion, the pressure drops again despite ongoing heating. When the heating ends, which is indicated by the circles, the temperature starts to drop as well and the systems dive through the binodal. Cavitation, however, sets in only when the systems cross the spinodal, which for the lower temperatures happens at significantly negative pressures. This process is very similar for all heating temperatures; what makes a difference for the highest temperatures is that here the spinodal is crossed at only slightly negative pressures. Therefore, even after expansion, the system stays close to the spinodal and rupturing continues until total disintegration is reached.
One can also observe that for the lowest simulated heat bath temperature (\SI{2000}{\kelvin}) the system does not reach the spinodal; in this case no void formation sets in. Instead the now liquid droplet exhibits a significant breathing mode vibration, a motion strongly damped by the void formation for the next three higher temperature cases which also end up in compact liquid droplets. Here, however, some breathing mode vibration is excited by the collapse of the cavity, most notably in the \SI{5000}{\kelvin} case.

In conclusion, we have demonstrated that the dynamics of strongly heated silver nanoparticles is dominated by a decompression wave caused by the near-isochoric heating, leading to distinct temperature and time dependent effects. 

Our results show that time-resolved x-ray diffraction experiments on nanoparticles in the gas phase are ideally suited to study superheated matter with high spatial and temporal resolution, thus providing access to material properties at very high temperatures and pressures.

\section{Acknowledgements}
We acknowledge DESY (Hamburg, Germany), a member of the Helmholtz Association HGF, for the provision of experimental facilities. 
Parts of this research were carried out at FLASH. Beamtime was allocated for proposal F-20170541.
This research was supported in part through the Maxwell computational resources operated at Deutsches Elektronen-Synchrotron DESY, Hamburg, Germany. We acknowledge the Max Planck Society for funding the development and the initial operation of the CAMP end-station within the Max Planck Advanced Study Group at CFEL and for providing this equipment for CAMP@FLASH. The installation of CAMP@FLASH was partially funded by the BMBF grants 05K10KT2, 05K13KT2, 05K16KT3 and 05K10KTB from FSP-302.
TM and BvI acknowledge funding by a DFG Koselleck Project MO 719/13 and IS61/14.
AC, LH, MS, and DR acknowledge funding form Leibniz society, Germany, via grant no. SAW/2017/MBI4 and from SNF, Switzerland, via grant no. 200021E/193642. TR and MM acknowledge the Gauss Centre for Supercomputing e.V.\ for providing computing time on the GCS Supercomputer JUWELS \cite{alvarez2021juwels} at Jülich Supercomputing Centre. Additional computing time was granted by the state of Baden-Württemberg through bwHPC and the DFG through grant no.\ INST39/961-1FUGG (bwForCluster NEMO).

\bibliographystyle{apsrev4-2}
\bibliography{paperBib}

\end{document}



\title{\Large{Supplemental Material for\\ Melting, bubble-like expansion and explosion of superheated plasmonic nanoparticles}}

\author{Simon Dold} 
\affiliation{\Freiburg}
\affiliation{\XFEL}
\author{Thomas Reichenbach} 
\affiliation{\Freiburg}
\affiliation{\IWM}
\author{Alessandro Colombo}
\affiliation{\ETH}
\author{Jakob Jordan}\affiliation{\TU}
\author{Ingo Barke}\affiliation{\Rostock}\affiliation{\RostockLLM}
\author{Patrick Behrens}\affiliation{\TU}
\author{Nils Bernhardt}\affiliation{\TU}
\author{Jonathan Correa}\affiliation{\FLASH}
\author{Stefan D\"usterer}\affiliation{\FLASH}
\author{Benjamin Erk}\affiliation{\FLASH}
\author{Thomas Fennel}\affiliation{\Rostock}\affiliation{\RostockLLM}
\author{Linos Hecht}\affiliation{\ETH}
\author{Andrea Heilrath}\affiliation{\TU}
\author{Robert Irsig}\affiliation{\Rostock}
\author{Norman Iwe}\affiliation{\Rostock}
\author{Patrice Kolb}\affiliation{\ETH}
\author{Bj\"orn Kruse}\affiliation{\Rostock}
\author{Bruno Langbehn}\affiliation{\TU}
\author{Bastian Manschwetus}\affiliation{\FLASH}
\author{Philipp Marienhagen}\affiliation{\RostockChem}
\author{Franklin Martinez}\affiliation{\Rostock}
\author{Karl-Heinz Meiwes-Broer}\affiliation{\Rostock}\affiliation{\RostockLLM}
\author{Kevin Oldenburg}\affiliation{\Rostock}\affiliation{\RostockLLM}
\author{Christopher Passow}\affiliation{\FLASH}
\author{Christian Peltz}\affiliation{\Rostock}
\author{Mario Sauppe}\affiliation{\TU}\affiliation{\ETH}
\author{Fabian Seel}\affiliation{\TU}
\author{Rico Mayro P. Tanyag}\affiliation{\TU}
\author{Rolf Treusch}\affiliation{\FLASH}
\author{Anatoli Ulmer}\affiliation{\TU}\affiliation{\Hamburg}
\author{Saida Walz}\affiliation{\TU}
\author{Michael Moseler} \affiliation{\Freiburg} \affiliation{\IWM}
\author{Thomas M\"oller}\affiliation{\TU}
\author{Daniela Rupp}
 \email{ruppda@phys.ethz.ch}
\affiliation{\ETH}\affiliation{\MBI}
\author{Bernd von Issendorff}
 \email{bernd.von.issendorff@uni-freiburg.de}
\affiliation{\Freiburg}\affiliation{\FMF}


\maketitle

\noindent \textbf{Contents}: Molecular dynamics simulations of strongly heated silver particles; Description of supplementary movies; Determination of onset of motion; Computation of phase diagram trajectories; Estimation of FEL pulse energies in the interaction region; Selection of small angle cutout for CDI; Further examples of diffraction patterns.

\newpage
\section{Molecular dynamics simulations of strongly heated silver particles}
To simulate the time evolution of strongly heated silver particles we used classical molecular dynamics (MD) simulations. Interatomic interactions were modelled using the embedded-atom-method potential of Sheng et al.~\cite{sheng2011highly}. As a starting configuration we cut spherical particles with 8024074 atoms from an fcc lattice of Ag atoms. After an initial relaxation, we subsequently heated the atoms for~10 ps using a Langevin thermostat with a time constant of 5~ps and a heat bath temperature between 2000~K and 9000~K. After the heating process the atomic positions were propagated in an NVE ensemble (constant particle number, volume and energy). Time integration was performed with a timestep of 0.1~fs.

We used LAMMPS \cite{thompson_lammps_2022} to perform the time integration, and ASE \cite{larsen_atomic_2017}, matscipy \cite{matscipy} and OVITO Pro~\cite{stukowski_visualization_2009} for visualization, pre- and postprocessing.

\section{Description of the supplementary movies}
\path{Sphere_8e6atoms_*K.mp4}:  Spherical Ag particles with 8024074 atoms that were coupled to a heat bath with a temperature between 2000~K and 9000~K for 10~ps.

\path{Sphere_8e6atoms_*K_cut.mp4}:  Cross section through the center of mass of the particles. Additionally, we used a surface construction \cite{stukowski_visualization_2009} with a probe sphere radius of 4~\AA\ to highlight the cavitations (shown in white).

\section{Determining the position of the boundary between
moving and non-moving layers during the heating process}
\begin{figure}
    \centering
    \includegraphics[width=\textwidth]{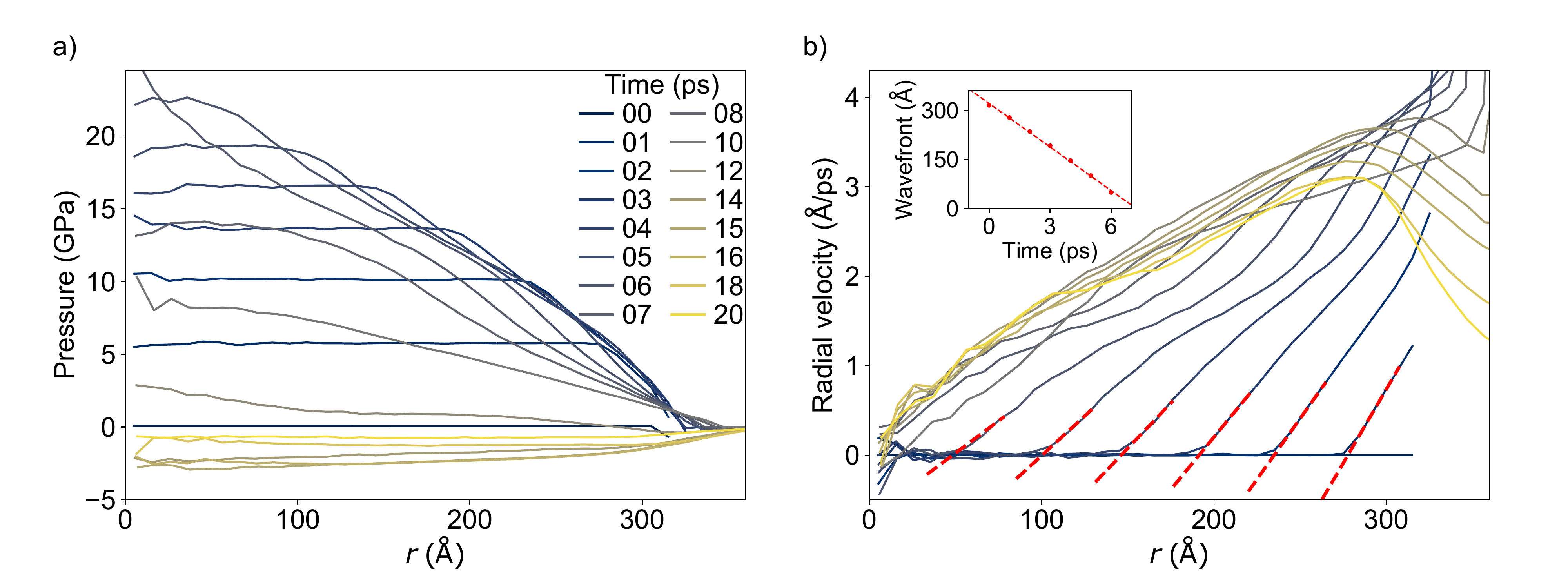}
    \caption{
    Radial pressure (a) and velocity (b) distributions at different times of a spherical silver particle with $8\times10^6$~atoms, heated for 10~ps by coupling to a heat bath with a temperature of $5000$~K. Averages are collected in radial bins of size 10~$\mathrm{\AA}$.
    In the calculation of the pressure, the average radial atomic velocity within each bin has been subtracted from the velocities of the atoms.
    In panel (b), the position of the boundary  between moving and non-moving material layers has been determined using the zero-crossing of the linear fits shown as  red dashed lines. The inset shows the position of the boundary as a function of time and the corresponding linear fit is what is shown in Fig.~3(a) in the main text. Its slope gives a velocity of sound of about 4400~m/s under these conditions.
    }
    \label{fig:fit_profiles}
\end{figure}

Figure \ref{fig:fit_profiles} shows the radial pressure and velocity distributions for a Ag particle that has been coupled to a heat bath with a temperature of 5000~K.
This data has been used to determine the radial position of the boundary between moving and non-moving layers in Fig.~3(a) in the main text. Details are given in the caption.

\section{Computation of phase diagram trajectories}
\subsection{Melting temperature as a function of hydrostatic pressure} To calculate the melting temperature of Ag for varying hydrostatic pressures between 0 and 15~GPa, we equilibrated a two-phase system consisting of a crystalline and a liquid phase in the NPH ensemble (constant particle number, pressure and enthalpy). The system is composed of 49392~atoms, has been set up in an orthogonal cell and is periodic along all three Cartesian directions. Two straight interfaces separate the two phases, both of which are initially equilibrated close to the melting temperature at the respective pressure. Upon equilibration in the NPH ensemble, the system temperature converges to the melting temperature as a result of crystal melting/crystallization processes. The pressure was controlled using a Parrinello-Rahman barostat~\cite{parrinello_polymorphic_1981}, where the lengths of the simulation cell vectors along the Cartesian directions were allowed to vary independently. Time integration was performed with a timestep of 2~fs.

\subsection{Binodal and spinodal}
To determine the binodal and spinodal of Ag, we calculated $P$-$V$ isotherms for temperatures between 1300~K and 5550~K (see e.g., Ref.~\cite{bogels_critical_2022} for a detailed discussion of this approach). For each temperature, we initially equilibrated 500125 Ag atoms in a cubic, periodic cell at the respective temperature and 1~GPa of hydrostatic pressure. Afterwards, we rescaled the atomic positions and the system’s volume to sample different densities. For each density, we equilibrated the system at constant volume and temperature for 300~ps, and measured the average hydrostatic pressure during the last 200~ps. The temperature was controlled using a Langevin thermostat with a time constant of 1~ps. Time integration was performed using a timestep of 1~fs. 

\begin{figure}
    \centering
    \includegraphics[width=0.6\textwidth]{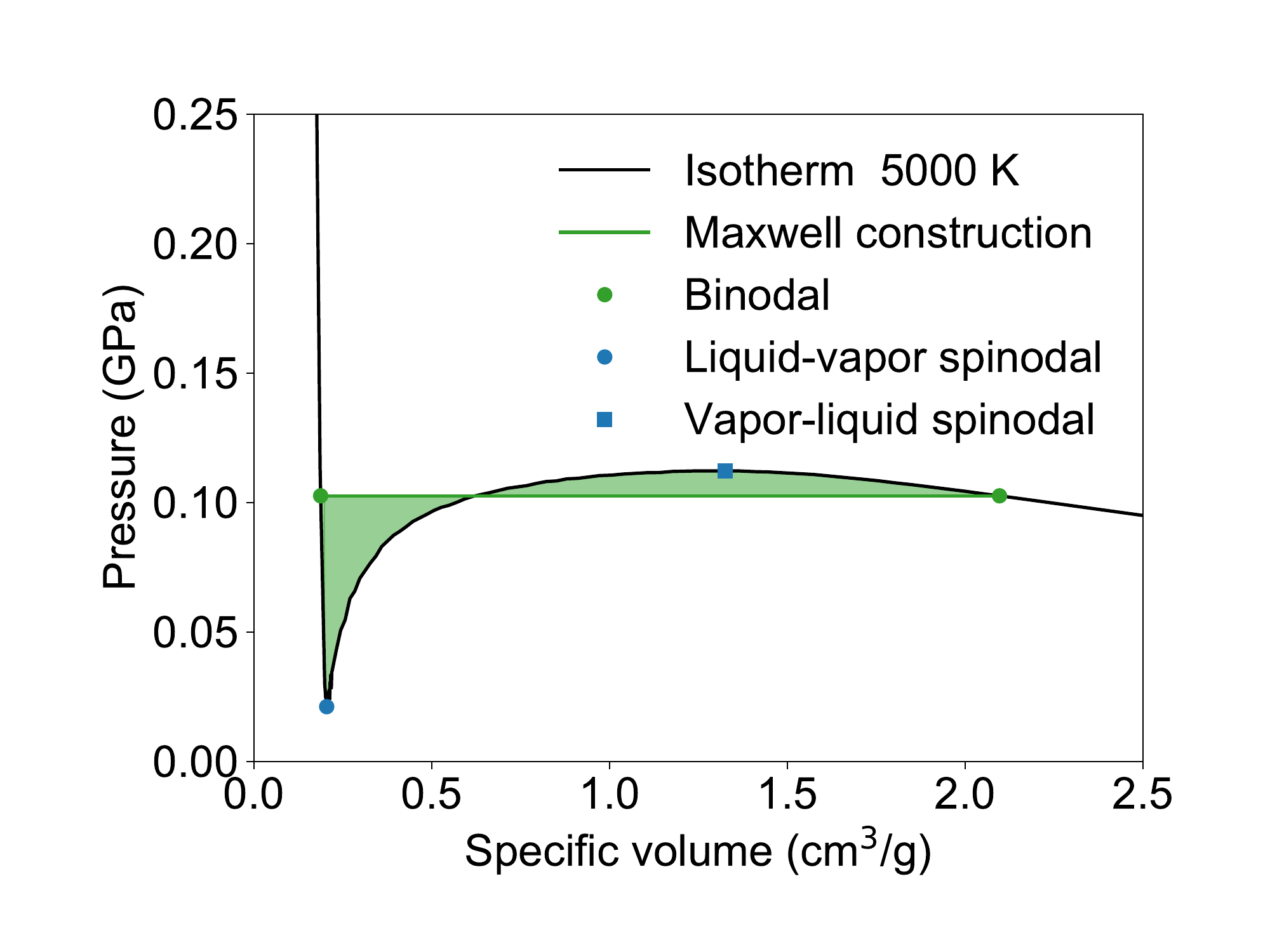}
    \caption{Pressure-volume isotherm of Ag at 5000~K. A Maxwell construction yields the binodal, the minimum and maximum correspond to the spinodals.}
    \label{fig:binode}
\end{figure}

From the resulting $P$-$V$ isotherms we determined the binode by Maxwell constructions (as exemplary illustrated in Fig.~\ref{fig:binode} for 5000~K). The extrema are the spinodals: the minimum corresponds to the liquid-vapor spinodal (shown in Fig.~3 in the main text), which limits the stability of the liquid phase; the maximum corresponds to the vapor-liquid spinodal, which limits the stability of the vapor phase. The critical point lies close to 5550~K and 0.18~GPa.
Since the binodal is very close to zero pressure at 1300~K, the solid-gas phase boundary in Fig.~3 in the main text is drawn as horizontal line with $P=0$.

\subsection{Phase diagram trajectories}
To estimate the trajectory of an Ag particle that has been coupled to a heat bath with a given temperature in the $P$-$T$ phase diagram, we calculated the radial density distribution for each snapshot of the system (every 1~ps after the onset of the heating process until 110~ps, every 5~ps afterwards). Averages were collected in radial bins with a width of 3~\AA. We defined the cutoff radius to determine the average pressure and temperature within the particles (while they are not yet ripped apart) as the center of the outermost radial bin, where the number density is larger than 0.01~\AA$^{-3}$. In the calculation of the average temperature and pressure within this cutoff radius, the average radial atomic velocity within each bin was subtracted from the velocities of the atoms.

\section{Estimate of FEL pulse energies in the interaction region}
Gas Monitor Detector (GMD) \cite{sorokin2019x} data gave an estimate of \SI{70}{\uJ} before the focusing mirrors. At \SI{5.1}{\nm}, the transmission of the mirrors is about \SI{27.8}{\percent} \cite{erk2018camp}. From this we can estimate \SI{19.5}{\uJ} in the interaction region.

\section{Selection of small angle cutout for CDI} 
CDI based on Iterative Phase Retrieval (IPR) algorithm relies on the assumption of small-angle scattering. CDI detectors count the number of photons scattered by an object for each pixel, which is proportional to the squared amplitude of the scattered field. When monochromatic light is employed, and the scattering is elastic, the scattered field lays on a spherical surface in reciprocal space, known as \emph{Ewald sphere}.
The small-angle approximation assumes that the maximum scattering angle at which light is recorded is sufficiently small that the portion of the \emph{Ewald sphere} accessed by the detector can be approximated as a flat surface. 
In this approximation, the scattered field is proportional to the square amplitude of the Fourier Transform of the object's electron density projected along the beam propagation direction, which is then reconstructed via IPR algorithms \cite{kirian2020imaging}.

The original diffraction data presented in this work cover scattering angles up to $\sim 30^\circ$, which is far from the small-angle approximation \cite{colombo2023three}. For this reason, only the inner part of the original scattering data can be effectively employed for imaging via IPR, for which the corresponding scattering angles are sufficiently small.

There is no strict definition of the threshold angle above which the small-angle approximation is no more valid, and scientific literature often indicates $10^\circ$. 
For the analysis presented here, diffraction data has been restricted up to a scattering angle of $15^\circ$ (half of the original maximum scattering angle) to maximize the spatial resolution of the reconstructions. It is worth noting that diffraction data between $10^\circ$ and $15^\circ$, which may be affected by the distortions of the \emph{Ewald sphere}, correspond to spatial resolutions close to the single pixel. This implies that the inclusion of these higher scattering angles in IPR can, in the worst case, only introduce artifacts at the length scale of single pixels. Reconstructed features and structures of two or more pixels are fully trustworthy, as their respective information in the reciprocal space is provided by scattering angles within $7.5^\circ$.


\section{Further information on the analysis of diffraction patterns}\label{diffinfo}


The scattering signal is considered up to a maximum transfer momentum
\begin{equation}
    q_\text{max} = \frac{4\pi}{\lambda} \sin(\frac{\theta_\text{max}}{2}) = \SI{0.32}{nm}^{-1}
    \end{equation}
with the cutoff scattering angle $\theta_\text{max} = 15^\circ$, and the radiation wavelength $\lambda = \SI{5.1}{nm}$. Under this condition, the reconstruction pixel size, also known as the \emph{half-period} resolution, is $\frac{\pi}{q_\text{max}} = \SI{9.8}{nm}$.

A total of 472 patterns were taken into account for the statistical analysis as function of the time delay, shown in Fig.~1. 
Only a subset of those patterns were actually suitable for analysis via IPR, which allows to obtain a real-space representation of the electron density of the objects. 
A diffraction pattern has to fulfill the following conditions to be suitable for a meaningful CDI via IPR algorithms:
\begin{enumerate}
    \item The diffraction data has to be sufficiently bright, with enough signal above the background noise to clearly identify the main features of the diffraction data.
    \item The size of the object has to be sufficiently large to have a corresponding size of the IPR reconstruction of more than few pixels, to allow a proper structural characterization
    \item The size of the object has to be sufficiently small to make the respective diffraction pattern satisfy the \emph{oversampling} condition \cite{kirian2020imaging}. In practical terms, this means that the amount of information contained in the diffraction pattern must be higher than the amount of information to retrieve (which is proportional to the object's size). For the dataset presented in this work, this condition is met for particles up to \SI{2.5}{\micro\meter} in diameter (which is half the pattern size in pixels, 256, times the pixel size, \SI{9.8}{\nano\meter}). However, this limit is in practice strongly reduced by the missing scattering information in the gap between the two detector modules (grey area in the patterns shown in Fig.~1 of the main text) and by the high level of noise. Thus, imaging via IPR was only successful for particle's sizes around or below \SI{250}{\nano\meter}.
\end{enumerate}

\begin{figure}
    \centering
    \includegraphics[width=0.34\textwidth]{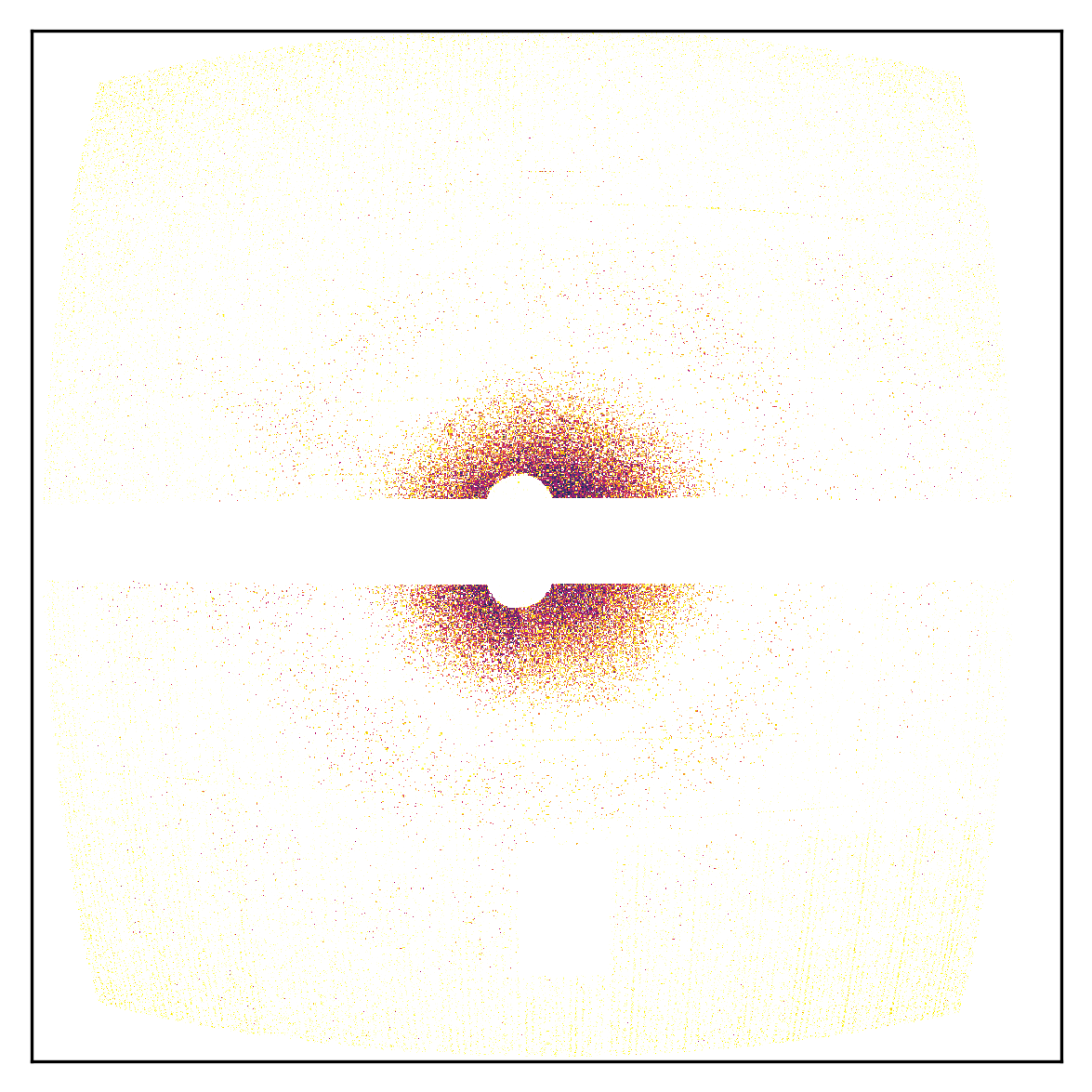}\includegraphics[width=0.34\textwidth]{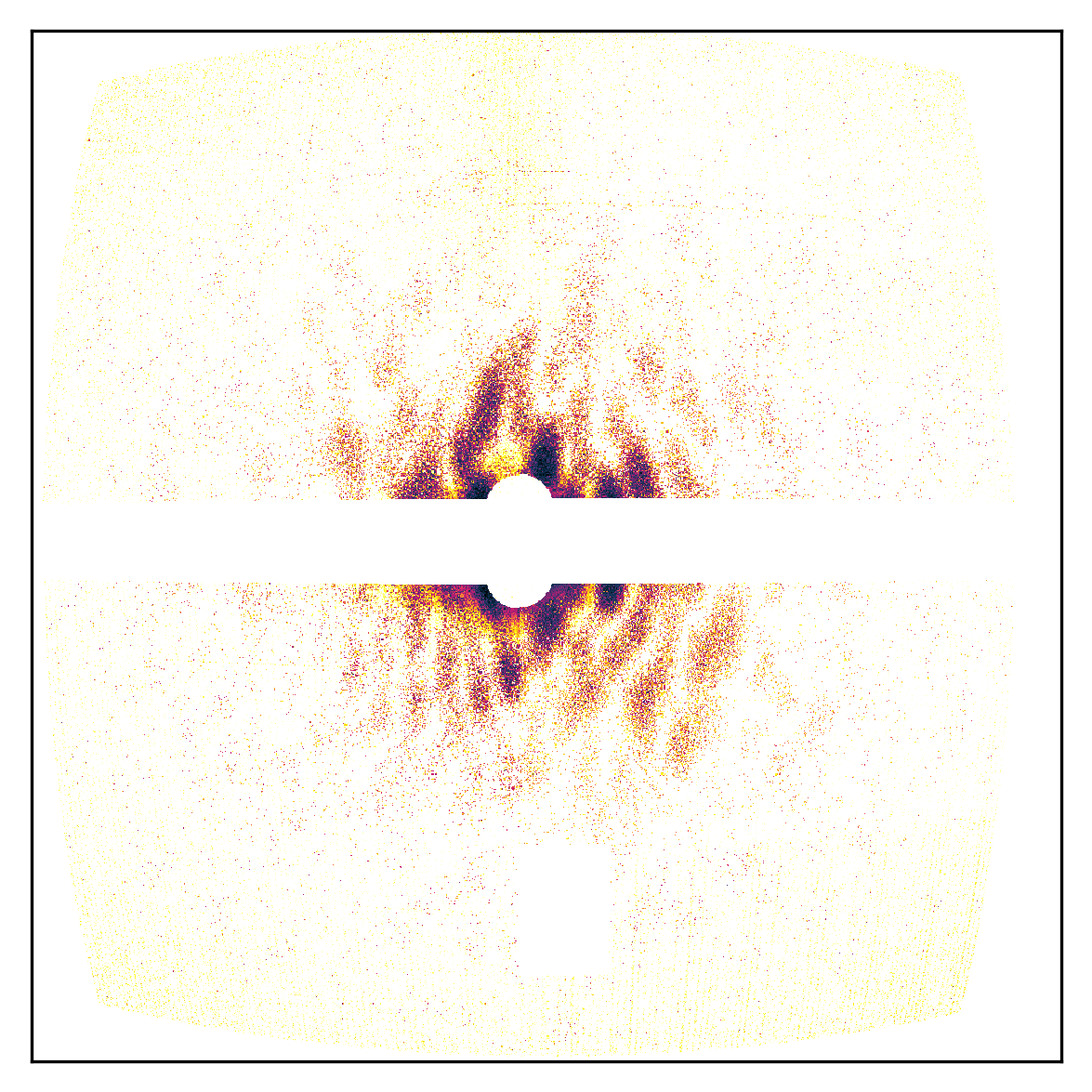}
    \caption{Examples of patterns that were excluded from the classification analysis. In these cases, the information provided by the diffraction data is not enough (left) or too complicated (right, corresponding to an agglomerate of particles) to effectively identify the structural properties of the objects. }
    \label{fig:excludedpatterns}
\end{figure}

Patterns not fulfilling conditions 1 and/or 2 have been completely excluded from the analysis, as the information provided by the scattering is too low for a successful classification. An example of such a pattern is shown in Fig.~\ref{fig:excludedpatterns}.

\begin{figure}
    \centering
    \includegraphics[width=0.9\textwidth]{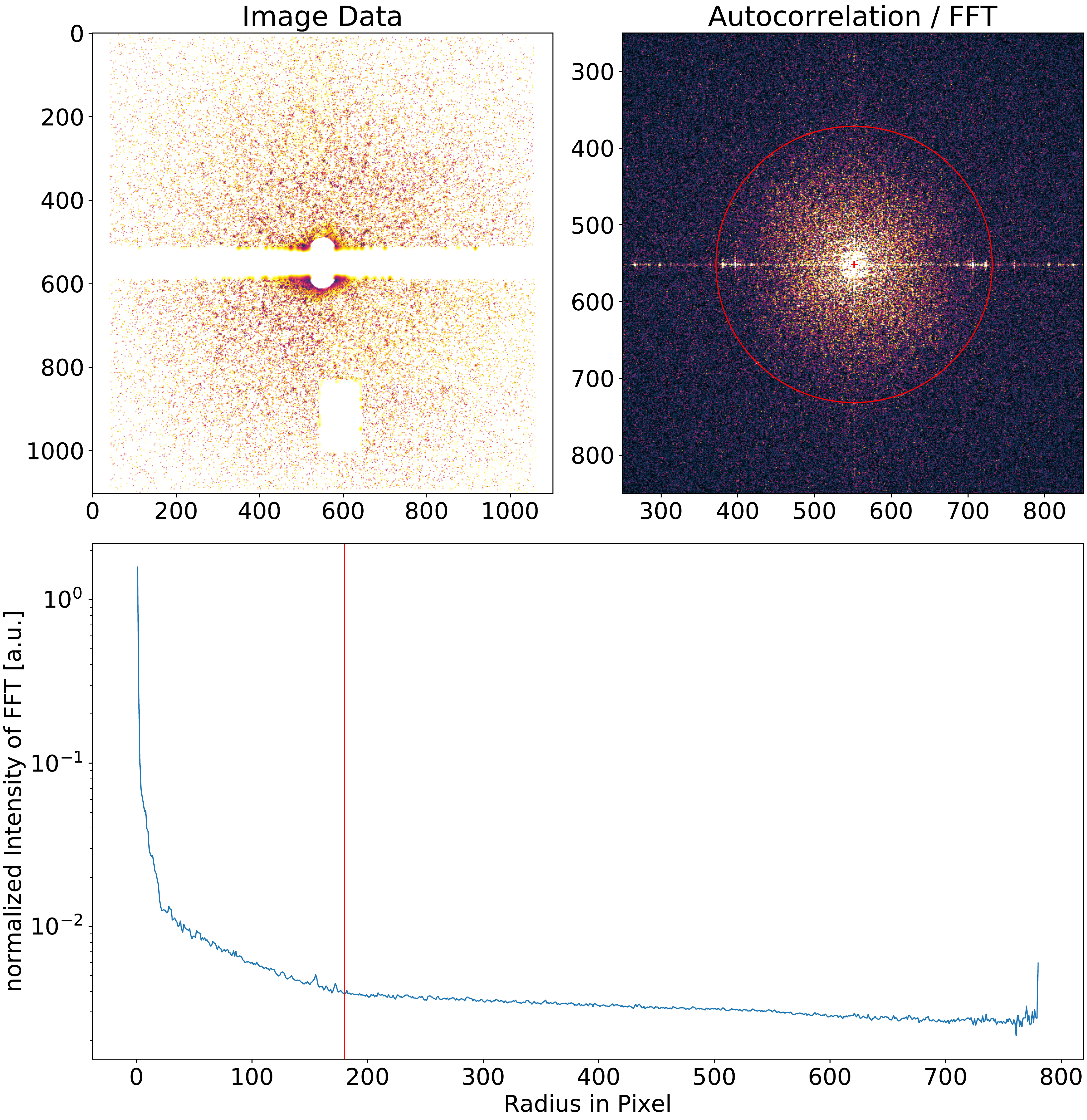}
    \caption{The spatial distribution cannot be recovered via IPR for many speckled patterns, due to a spatial extension of the sample that does not fulfill the oversampling condition. However, comparison with other patterns as well as an analysis of the sample's autocorrelation function (obtained by performing a FT of the diffraction pattern) still allow their size evaluation. The extend of an approximately spherical symmetrically scatterer is given by half the extend (red line) of the radial profile of the autocorrelation (lower panel). Pixel size about \SI{5}{\nm}.}
    \label{fig:bigpatterns}
\end{figure}

A different approach is undertaken for those patterns that do not fulfill condition 3 only. In this case, enough information on the object is contained in the experimental data, but the imaging via IPR fails to converge to a meaningful solution. Structural properties of the object can be deduced by comparing features of the diffraction pattern of interest with those of a diffraction data that was successfully imaged via IPR. 
In most cases, those patterns are produced by exploded silver clusters. The cluster's fragments, due to their kinetic energy, spread out in space, immensely increasing the overall spatial extension of the sample. However, it is still possible to deduce their real-space properties as shown in Fig.~\ref{fig:bigpatterns}. The real space sizes of the pixels can be calculated as described in \ref{diffinfo}.

\begin{figure}
    \centering
    \includegraphics[width=0.6\textwidth]{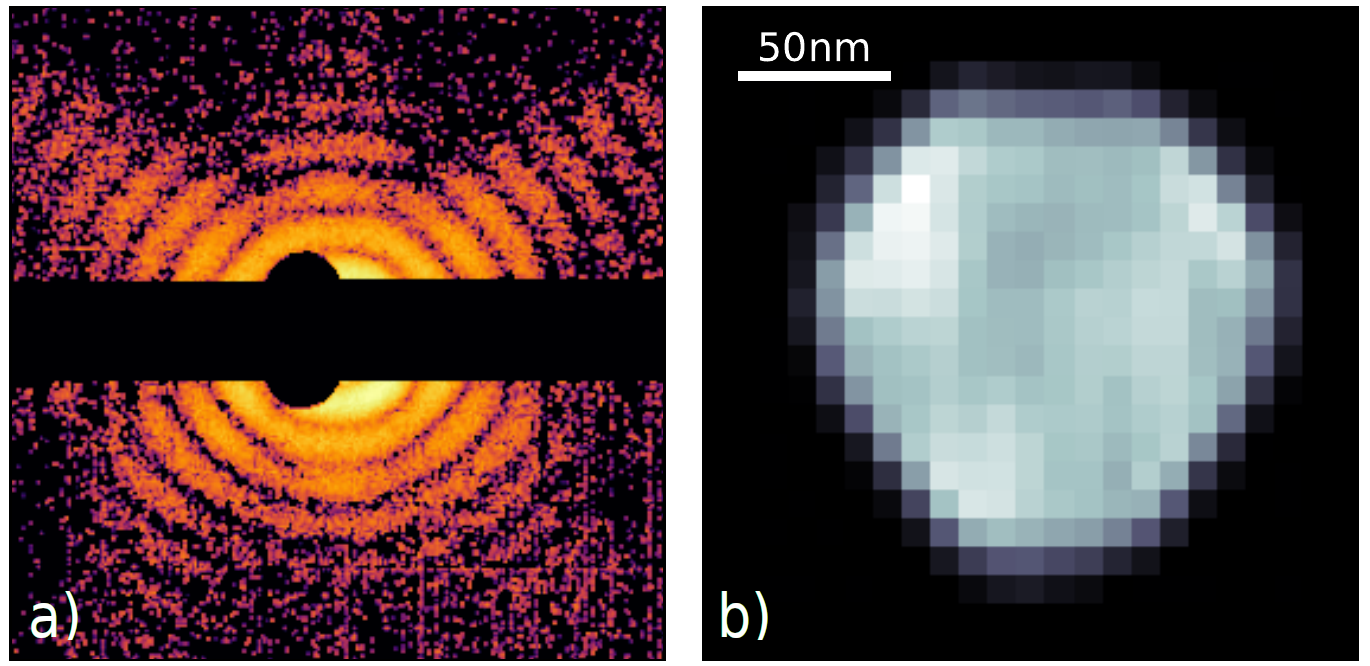}
    \caption{Example of an ``uncertain'' diffraction pattern. The small-angle experimental data is shown in a), while the reconstruction, displaying the object's electronic density, is shown in b). The electron density in the inner part suggests the presence of a low-density region, whose magnitude is, however, comparable with artifacts produced by noise.}
    \label{fig:uncertain}
\end{figure}

Among those patterns that contain enough information, some of them were labeled as ``uncertain''. This means that their diffraction patterns (or their respective IPR reconstructions) present some features that are, however, not strong enough to assign them a category label with reasonable certainty, as exemplified in Fig.~\ref{fig:uncertain}. These diffraction shots have been included in the total count of diffraction patterns, but were not analyzed further. A similar consideration applies for those patterns produced by clusters agglomerates \cite{colombo2023three}.

\FloatBarrier
\bibliographystyle{apsrev4-2}
\bibliography{paperBib}